\documentclass[12pt]{article}
\usepackage{vmargin}
\usepackage{xspace}
\setmarginsrb{2.4cm}{2cm}{2.4cm}{2.5cm}{1.5cm}{0cm}{.5cm}{1.4cm}

\newcommand{\comment}[1]{}
\newcommand{\cR}{{\mathcal R}}

\usepackage{pifont,amssymb}
\newcommand{\es}{\mbox{\Pisymbol{psy}{198}}}
\newcommand{\GSt}{{\sc GST[1,2]\xspace}}
\newcommand{\St}{{\sc STP[1,2]\xspace}}
\newcommand{\Tr}{T_{\it ref}}
\newcommand{\Tsk}{T_{\it ref}^{\it sk}}
\newcommand{\twth}{\frac{2}{3}}
\newcommand{\onth}{\frac{1}{3}}

\newcommand{\onsi}{\frac{1}{6}}
\newcommand{\half}{\frac{1}{2}}
\newcommand{\thha}{\frac{3}{2}}
\newcommand{\Prom}{{\it PromCost}\xspace}
\newcommand{\prom}{\mbox{\Prom}}
\newenvironment{prf}{\noindent{\bf Proof.}\ }{\hfill\ding{114}\smallskip}
\newtheorem{lemma}{Lemma}
\newtheorem{theorem}{Theorem}
\title{\bf A Factor 3/2 Approximation for\\ Generalized Steiner Tree Problem with\\
Distances One and Two\vspace{0.4cm}}
\author{Piotr Berman\thanks{Department of Computer Science \& Engineering,
Pennsylvania State University, University Park, PA 16802. Research partially
done while visiting Department of Computer Science, University of Bonn and
supported by DFG grant Bo 56/174-1. Email: berman@cse.psu.edu}
\and Marek Karpinski\thanks{Department of Computer Science and the Hausdorff Center for Mathematics, University of
Bonn, 53117 Bonn. Supported in part by DFG grants, Procope grant 31022,
and the Hausdorff Center grant EXC59-1. Email: marek@cs.uni-bonn.de}
\and Alex Zelikovsky\thanks{Department of Computer Science, Georgia State
University, Atlanta, GA 30303. Email: alexz@cs.gsu.edu}
}
\date{}

\begin{document}
\maketitle
\abstract{We design a 3/2 approximation algorithm for  the Generalized
Steiner Tree problem (GST) in metrics with distances 1 and 2.
This is the first polynomial time approximation algorithm for a wide class of non-geometric metric GST
instances with approximation factor below 2.\\[1ex]
}
\section{Introduction}
We design a 3/2 approximation algorithm for constructing
generalized Steiner trees (Steiner Forests) for metrics with distances 1 and 2.
With the exception of geometric metrics \cite{BKM}, there were no wide classes
of instances known with approximation ratios better than 2.  This was
in contrast to similar problems like Traveling Salesman
and Steiner Tree Problems \cite{BK}, \cite{BKZ}.

\section{Definitions and Notation}

A metric with distances 1 and 2 can be represented as a graph with edges being
pairs of distance 1 and non-edges being pairs of distance 2.
We will use GST[1,2] to denote the Generalized Steiner Tree Problem restricted
to such metrics.

The problem instance of GST(1,2)
will be a graph $G=(V,E)$ that defines a metric in this way, and a collection $\cR$
of subsets of $V$ called {\em required sets}.
We say that $\cup_{R\in\cR}R$ is the set of {\em terminals}.
In a {\em proper} instance, the required sets do not overlap and have more than
one element.
It is obvious that for every family of requirements
$\cR$, there exists a unique family $prop(\cR)$ that is equivalent and proper.

A valid solution is a set of unordered node pairs $F$ such that each $R_i$ is
contained in a connected component of $(V,F)$. The objective is to minimize $|F\cap E|+2|F-E|$.

We will use in the sequel some notation and terminology introduced in \cite{BKZ}.

A basic building block of our solutions is an $s$-star consisting of a non-terminal $c$, called the center, $s$ terminals
$t_1,\ldots,t_s$ and edges $(c,t_1),\ldots,(c,t_s)$. In \cite{BKZ} we used also a more general version of a building block, an $(r,s)$-comet consisting of a non-terminal center $c$, non-terminal fork nodes $f_1,\ldots,f_s$ plus $r+2s$ terminals, the center is connected to $r$ terminals and all the fork nodes, while each fork node is connected to two terminals of its own.

If $s<3$ we say that the star is {\em degenerate}, and {\em proper}, otherwise.

In the analysis of our algorithm, we will view its selections as transformations of an
input instance, so after each phase we have a partial solution and a residual
instance.  We formalize these notions as follows.

A partition $\Pi$ of $V$ induces a graph $(\Pi,E(\Pi))$ where
$(A,B)\in E(\Pi)$ if $(u,v)\in E$ for some $u\in A,v\in B)$.
We say that $(u,v)$ is a representative of $(A,B)$.

Similarly, $\Pi$ induces required sets.  Let
$R_\Pi=\{A\in\Pi:A\cap R\not=\es\}$, then $\cR_\Pi=prop(\{R_\Pi:~R\in\cR\})$.

In our algorithms, we augment initially empty solution $F$.
Edge set $F$ defines partition $\Pi(F)$ into connected components of $(V,F)$.
In a step, we identify a connected set $A$ in the induced graph
$(\Pi(F),E(\Pi(F)))$ and we augment $F$ with representatives of edges
that form a spanning tree of $A$.  We will call it ``collapsing $A$'',
because $A$ will become a single node of $(\Pi(F),E(\Pi(F)))$.

Thus, if we select some "building block" $C$, $F$ is going to be augmented by the representatives of the edges in $C$, and this changes the "residual" graph in which we make our next selection. For that reason we will use terms "select" and
"collapse" as synonyms,

\section{Analyzing Greedy Heuristics}
\label{ARSh}

We introduce a new way of analyzing greedy heuristics for our
problem, and in this section we illustrate it on the example of the Rayward-Smith
heuristic \cite{R-S} for \St. This heuristic has approximation ratio of
exactly $4/3$, as demonstrated by Bern and Plassman \cite{BP}.
However, the new analysis method is tighter
(see Theorem \ref{fouroverthree}) and characterizes the
effect of more general classes of greedy choices, as we will show
in the next section. We have reformulated the Rayward-Smith
heuristic as follows.\\[1ex]

\fbox{
  \parbox[c]{4.7in}{
\medskip

{\bf \ While} there is more than one terminal

{\ \ \ \ perform} the first possible operation from the following list:
\begin{description}
\item[{\bf \ \ \ \ 1. Preprocessing:}] Collapse an edge between terminals.
\item[{\bf \ \ \ \ 2. Collapsing of stars:}] Collapse an $s$-star $S$ with maximum $s$.
\item[{\bf \ \ \ \ 3. Finishing:}] Connect two terminals with a non-edge.
\end{description}
}
}\\[1ex]



If we can perform a step of Preprocessing, the
approximation ratio can only improve
since such the collapsed edge can be forced into the
optimal solution.  Thus it suffices to analyze the case when
no two terminals share a cost-1 connection.

Let $T^*$ be an optimal Steiner tree  and let $T=T^* \cap E$ be its
{\em Steiner skeleton} consisting of its edges (cost-1 connections),

Let $T_{RS}$ be the Steiner tree given by Rayward-Smith heuristic.
We are going to prove the following

\begin{theorem}
\label{fouroverthree} $cost(T_{RS}) < cost(T^*)+ {1\over 3} cost(T)$.
\end{theorem}

In the analysis of the Collapsing of stars and Finishing, we
update the following three values after each iteration:

\begin{itemize}
\item[$CA$]
 = the total cost $F$, the set of edges  collapsed so far, initially, $CA=0$;

\item[$CR$]
 = the cost of the {\em reference solution} $\Tr$ derived from
 the optimum solution $T^*$; $\Tr$ is a solution of the
 residual problem in $(V_F,E_F)$;

\item[$P$]
 = the sum of potentials distributed among objects, which will be
 defined later.
\end{itemize}

The sum $CA+CR+P$ will be the {\it promised cost}, \Prom.

We will define the potential satisfying the following conditions:

\begin{itemize}
\item[(a)] initially, $P< cost(T)/3\le cost(T^*)/3$;
\item[(b)] after each star collapse, \Prom will be unchanged
or decreased.
\item[(c)] at termination, $CR=0$ ($\Tr$ will be empty) and $P=0$.
\end{itemize}

These properties clearly imply our claim, as the initial \Prom
would satisfy the statement of the theorem, \Prom cannot increase
and at the termination we return a solution with that cost.

Initially, $\Tr=T=T^*\cap E$.
In the analysis, we also use the {\em skeleton} of $\Tr$,
$\Tsk=\Tr\cap E$, the set of 1-cost connections of $\Tr$.
The potential is given to the following objects:
\begin{itemize}
\item
edges of $\Tsk$;
\item
{\em C-comps} which are connected components of $\Tsk$;
\item
{\em S-comps} which are Steiner full components of $\Tsk$.
\end{itemize}
The total
potential of edges, C-comps and S-comps is denoted $PE$, $PC$ and $PS$
respectively.
{\bf At all times}, the potential of each edge $e\in\Tsk$ is $p(e)=\onth$.

Initially, the potential of each C-comp and S-comp is zero.

A Steiner tree is called {\em bridgeless} if no two Steiner points
are adjacent and each Steiner point has degree at least 3.

\begin{lemma} \label{Prep}
Without increasing \Prom, we can transform the optimum solution $T^*$ into a
bridgeless reference solution $\Tr$, while the new potential $p$ satisfies
\begin{itemize}
\item[(i)] each C-comp $C$ has $p(C)\ge -\twth$ and if $C$ has fewer than
3 edges, $p(C)=0$;
\item [(ii)] each S-comp $S$ has $p(S)=0$;
\end{itemize}
\end{lemma}

\begin{prf}
Because $CA$ and $PS$ remain zero, to
see that \Prom does not increase it suffices that each transformation
of $\Tr$ and $p$ satisfies
$\Delta CR+\Delta PE+\Delta PC\le 0$.
The bridgeless Steiner tree is obtained using
the following two types of steps.

\noindent {\bf Path step.} Suppose that $\Tsk$ contains a
Steiner point $v$ of degree 2. We remove two edges incident to $v$
from $\Tsk$ adding a non-edge (cost-2 connection) to $\Tr$. The
potential for the both resulting C-comps is set to 0.
One can see that $\Delta CR=0$, $\Delta PE=-\twth$ (two edges removed)
and $\Delta PC\le \twth$ ($\Delta PC=\twth$ if the component $C$ that
was split had $p(C)=-\twth$.

If the removal of edges in a Path step creates Steiner points of degree 1,
we remove them; this can only decrease \Prom.

\noindent{\bf Bridge Step.} Suppose that we cannot perform a Path step
and $e\in\Tsk$ is a {\em bridge}, i.e., an edge $e=(u,v)$
between Steiner points. We remove this edge from $\Tr$ (replacing with a
non-edge between terminals); this splits a C-comp $C$ into
$C_0$ and $C_1$. Each new C-comp has at lest two edges since $u$ and
$v$ originally have degree at least 3. We set $p(C_0)=p(C)$ and
$p(C_1)=-\twth$.  Thus $\Delta CR=1$ (the cost is increased by 1),
$\Delta PE=-\onth$ and $\Delta PC=-\twth$ (one more C-comp with potential
$-\twth$.

Note that if we create a C-comp with two edges, we can apply a Bridge Step;
this is because we assume than there are no edges between terminals.
\end{prf}

From now on our reference Steiner tree $\Tr$ is assumed to be bridgeless.

\noindent Now we will prove

\begin{lemma} \label{afterPrep}
After collapsing an $s$-star $S$, $s>3$, conditions (i)-(ii) of
Lemma~\ref{Prep} are satisfied and \Prom does not increase.
\end{lemma}

\begin{prf}
Suppose that the terminals of $S$ be in $a$ C-comps. To break cycles created
in $\Tr$ when we collapse $S$, we replace $s-1$ connections, of which $a-1$
are cost-2 connections between different C-comps and $s-a$ edges within
C-comps.

If this is the entire modification,
$\Delta CA=s$, $\Delta CR=-s-a+2$, $\Delta PE=-\onth(s-a)$ (for edges
removed from $\Tsk$) while $\Delta PC\le \twth(a-1)$ (for removing potential
of $a-1$ C-comps, each $-\twth$ or 0) hence

\centerline{$\Delta\prom=s -s-a+2
-\onth(s-a)+\twth(a-1)=\onth(4-s)\le 0.$}

However, the new C-comp that we create can be trivial; in this case we
need to increase the estimate of $\Delta PC$ by $\twth$.  If that C-comp had
but one edge left, this edge would be removed from $\Tr$ and $\Tsk$, which
decreases the estimate of $\Delta CR$ by 1 and $\Delta CE$ by $\onth$.
If that C-comp had two edges left, we would remove them from $\Tsk$
using a Path step, this does not change CR but decreases CE by $\twth$.
Therefore our estimate of \Prom does not increase.
\end{prf}

Once we collapsed $s$-stars for $s>3$ we
redistribute potential between C-comps and S-comps by
increasing potential of each nontrivial C-comp by $1\over 6$
bringing it to $-{1\over 2}$ and decreasing potential of one of its
S-comps to $-{1\over 6}$. This will replace conditions (i)-(ii) with
\begin{itemize}
\item[(i')] each C-comp $C$ has $p(C)\ge -\half$ and each trivial
C-comp (with at most one edge) has $p(C)=0$;
\item [(ii')] each S-comp $S$ has $p(S)\ge -{1\over 6}$;
\end{itemize}

\begin{lemma} \label{threestar}
After collapsing a 3-star, conditions (i')-(ii')
are satisfied and \Prom does not increase.
\end{lemma}

\begin{prf}
Suppose that the terminals of the selected star $S$ belong to 3 different
C-comps.  Then we replace two cost-2
connections from $\Tr$ with 3 collapsed edges, while we decrease the
number of C-comps by 2, thus

\centerline{$\Delta\prom=\Delta CA+\Delta CR+\Delta PC \le
3-4+2\frac{1}{2}=0$.}

Suppose that the terminals of $S$ belong to 2 different C-comps.
$\Delta CR=3$ because we remove one cost-2
connection from $\Tr$ and one edge from an S-comp.  This S-comp
becomes a 2-star, hence we remove it from $T$ using a Path Step, so
together we remove 3 edges from $\Tsk$ and $\Delta PE=1$.

One S-comp disappears, so $\Delta PS=-\onsi$. Because we collapse two
C-comps into one, $\Delta PC= -\half$.  Consequently,

\centerline{$\Delta\prom=
3-3-3\frac{1}{3}+\frac{1}{2}+\frac{1}{6}<0.$}

If the terminals of the selected star belong to a single C-comp
and we remove 2 edges from a single S-comp, we also remove the third edge of this S-comp and
$\Delta CR=-3$, while $\Delta PE=-1$, $\Delta PS=\onsi$, and if its C-comp
degenerates to a single node, we have $\Delta PC=\half$ (otherwise, zero).
This yields the same change in \Prom as the previous case.

Finally, if the terminals of the selected star belong to a single C-comp and
we remove 2 edges from two S-comps, we have $\Delta CA+\Delta CR= 1$.
Because we apply
Path Steps to those two S-comps, $\Delta PE=-2$.  while $\Delta PS=\onth$ and
$\Delta PC\le \half$.  Thus $\Delta\prom$ is at most $-\onsi$.
\end{prf}

To complete the proof of Theorem \ref{fouroverthree}
it suffices to see that when no more star collapsing is possible,
$\Tr$ consists of cost 2-connections, $\Tsk=\es$ and thus the
remaining potential is zero.  Each finishing step increases $CA$ by
2 and decreases $CR$ by 2, with no changes in \Prom.  When we
terminate, we have a solution with cost $CA=\prom$.


\section{3/2 Approximation for GST with Distances 1 and 2}
\label{generalized}

In the heuristic for \St we could start with {\bf Preprocessing} in which
we collapsed every edge (cost-1 connection) between terminals, arguing
that such an edge can be forced as a component of an optimum solution $T^*$.
In \GSt this is no longer valid, because this could be an edge
between different connected components of $T^*$.  Indeed, we need to
increase our potential and thus \Prom to create a ``budget'' for this class
of wrong selections: connecting sets that should not be connected.

Instead, we can start with the following preprocessing that is safe
in the context of \GSt:\\[1ex]

\fbox{
{\ \ }\parbox[c]{4.8in}{
\medskip
{\bf G-Preprocessing:} While there exists an edge or
an $s$-star (with $s\ge 3$)\\ contained in one of the required sets $R_i$,
collapse it.
\medskip
}
}\\[1ex]

We can also normalize the optimum solution $T^*$ to assure these two
properties: Steiner nodes have degree at least 3 and cost-2 connections
connect only pairs of terminals from the same required set $R_i$.
Steiner nodes of degree 1 can be obviously removed, Steiner nodes of degree 2
and cost-2 connections can be removed, and reconnection, if needed, can
be achieved by connecting terminals that have to be connected.

Because the terminals of the edge ($s$-star) selected by G-Preprocessing
is surely contained in a single connected component of the optimum forest
$T^*$, while we increase $CA$ by 1 ($s$), we decrease the cost of the
$T^*$ of the residual problem by 1 ($s-1$), thus preserving the approximation
ratio of 1/1 ($s/(s-1)$).

Thus we can proceed with the assumption that no steps of G-Preprocessing
can be performed.
After the preprocessing, we can perform normalization
of the ``reference tree'' $\Tr$ that we initialize with $T^*$.
Because $\Tr$ has multiple connected component and it may also contain
edges between terminals, we introduce two new notions:
\begin{itemize}
\item
{\em F-comps} which are connected components of the forest $\Tr$;
\item
{\em T-comps} which are connected components of the subgraph of
$\Tsk$ that is induced by the terminals.
\end{itemize}

We also introduce the second component of the potential, $p_g$,
such that the sum of all $p$(object) and $p_g$(object) does not
exceed $\half cost(T^*)$.  We will use $p_g$ to cover the cost of
connections made between different F-comps, the class of errors that are
specific to the generalized problem.

We can give $p_g(e)=\onsi$ for every edge of $T^*\cap E$,
for edges inside a $T$-comp we can increase it to $p_g(e)=\half$.
For each non-edge $e'$ in $T^*$ we can give $p_g(e')=1$.
Moreover, to each initial C-comp $C$ we can give $p_g(C)=-\twth$.
Let $p_g(F)$ be the sum of $p_g$ potentials of objects contained
in an F-comp $F$.

We can define $\prom'=\prom+PF$ where $PF$ is the sum of all $p_g(F)$'s.
Our goal is to build a solution by collapsing selected connections without
increasing $\prom'$.  When we make a connection within an F-comp,
we do not increase \Prom and $PF$ does not increase either.

When we make a selection that connects two F-comps, say $F_1,F_2$ into
$F=F_1\cup F_2$, we can
cover the cost of that connection using $p_g(F_1)$, and $F$ can use
$p_g(F_2)$ for a future connection with another F-comp.
Because we will not connect distinct F-comps with non-edges, such a
connection costs at most $\thha$ (this is the cost of connections
made by a 3-stars, larger stars and edges make connections with a smaller
cost).  This is safe if $p_g(F_i)\ge\thha$.

\begin{lemma}
If a required set of terminals $R_i$ has more than 2 nodes, it is contained
in an initial F-comp $F$ such that $p_g(F)\ge\thha$.
\end{lemma}
\begin{prf}
Tree $F$ may contain three kinds of connections: $e$ is a T-connection if
it is an edge between terminals, and $p_g(e)=\half$, a 2-connection if
it is a non-edge, and $p_g(e)=1$ and a C-connection, any other edge, and
$p_g(e)=\onsi$.

If $F$ contains a C-connection, it contains a Steiner node, and thus at least
3 C-connections and a C-comp; those objects alone give $p_g$ of
$3\onsi+\twth=\thha-\onsi$.  If there are no other connections in $F$, it
is a 3-star, but in this case all terminals of that star are in $R_i$
and we would collapse it in G-Preprocessing.  And
any other connection would increase $p_g(F)$ to at least $\thha$.

Now we assume that $F$ does not contain C-connections.  If it contains
a 2-connection, it must contain another connection as well, and the least
possible $p_g(F)$ is $1+\half$ if this other connection is a T-connection.
In the remaining case, $F$ has some $a$ terminals, $a-1$ T-connections
and $p_g(F)=(a-1)\half$.  Again, if $a>3$ then $p_g(F)$ is sufficiently
high and if $a=3$ then only $R_i$ are terminals of $F$ and the T-connections
would be collapsed in G-preprocessing.
\end{prf}

We can also observe that
\begin{lemma}
If a required set of terminals $R_i$ has 2 nodes, it is contained
in an initial F-comp $F$ such that $p_g(F)\ge 1$.
\end{lemma}

For this reason, it always safe to collapse edges between terminals.  However,
the status of the resulting merged sets of terminals requires some reasoning.
Let us say that a set of terminals $F$ is {\em safe} if it has $p_g(F)\ge\thha$.
When we merge two sets of terminals, $F_1$ and $F_2$ using a connection with
cost $c$, the union $F=F_1\cup F_2$ will get $p_g(F)=p_g(F_1)+p_g(F_2)-c$.
If $c=1$, then union is safe as long as at least one of $F_1,F_2$ is safe,
but not otherwise.  However, suppose that after a union creating a larger
unsafe set $F$ an edge (of the residual graph) is contained in $F$.
Then the balance of $F$ is more favorable, by 1, then our pessimistic
reasoning that deemed $F$ unsafe, and this suffices to tag is safe.

Thus we can perform a bit bolder preprocessing if we
keep track which resulting requirement sets are safe, and which are not.\\[1ex]

\fbox{
{\ \ }\parbox[c]{4.8in}{
\medskip
{\bf GE-Preprocessing} (G-preprocessing, extended version):
Tag each required set
of terminals $R_i$ as safe if $|R_i|>2$ and unsafe otherwise.
While you can, do the following: collapse $s$-star contained in
a required set and collapse any edge between two terminals.  In the
latter case, if these two terminals were in two different required sets
and thus the collapsing replaces them with their union,
tag the union safe if at least one of the merged requirement was safe.
Moreover, if the collapsed edge is contained in some requirement set,
tag that set safe.
\medskip
}
{\ \ }}\\[1ex]

We are now left with the problem: what to do with the unsafe sets that
remain after the GE-Preprocessing.  To address this problem, we
need a stronger version of Lemma \ref{Prep}.
\begin{lemma}
Without increasing \Prom we can transform the optimum solution
$T^*$ into a reference solution that satisfies the conditions of
Lemma \ref{Prep} and in which each
T-comp has a cost-1 connection to at most one Steiner point.
\end{lemma}
\begin{prf}
The reasoning is the same as in Lemma \ref{Prep}, except that we need to perform
Bridge Step in the situation when we have a T-comp connected to more
then one Steiner node.  If we cannot remove such a connection as a Path step,
we break a C-comp into two, so each of the resulting parts has at least
two edges adjacent to Steiner nodes (the sufficient premise for reasoning
of the Bridge Step).
\end{prf}

Now we can justify\\[1ex]

\fbox{
{\ \ }\parbox[c]{4.8in}{
\medskip
{\bf Annihilation of unsafe sets:}  After GE-Preprocessing,
break each unsafe set of requirements into original requirements,
connect them individually with cost-2 connections and remove
from further consideration.\\[-1.2ex]
}
{\ \ }}\\[1ex]

\begin{lemma}
{\bf Annihilation of unsafe sets} does not increase $\prom'$.
\end{lemma}
\begin{prf}
Consider an unsafe requirement $R'$.  It is created from a
union of some $p$ pair
requirements $R_1,\ldots,R_p$ (each with two terminals).
Because $R'$ remains unsafe, GE-preprocessing did not collapsed
exactly $p-1$ 1-cost connections inside $R'$, so it consists of
$p+1$ connected components (T-comps).
We increase $CA$ by $p+1$ by replacing these $p-1$ connections with
$p$ non-edges.

A pair $R_i$ that is connected separately in $\Tr$ contributes 3 to
$\prom'$ and after annihilation uses only the correct cost, 2, so this
case has a surplus.  One can see that the most tight case is when
in $\Tr$ every T-comp of $R'$ is connected by an edge to some Steiner
node (a connection to a terminal would be performed already).  Thus
we remove those connections and decrease $CR$ by $p+1$, remove the
connections made by GE-Preprocessing and decrease $CA$ by $p-1$ and
reconnect with $p$ non-edges; this does not change $CA+CR$, while the
sum of potentials can only decrease.
\end{prf}

We conclude that after the Annihilation of unsafe
sets we can proceed with the heuristic described in the previous section
without increasing $\prom'$, and $\prom'$ is initialized as not larger
than $\thha cost(T^*)$.

We construct now our approximation algorithm to consist of  GE-Preprocessing
followed by Annihilation of unsafe sets and followed by Rayward-Smith heuristics.

With the above we have the following main result.

\begin{theorem}
There exists a polynomial time 3/2-approximation algorithm
 for the
\GSt.
\end{theorem}

\end{document}